# Superconducting Flux Memory for Cryogenic Applications

Northrop Grumman Microelectronics Design and Applications*

Northrop Grumman Systems Corporation

9020 Junction Dr, Annapolis Junction, MD, 20701

## Abstract

We report the development of flux memory for use with superconducting circuits. This technology stores persistent currents in superconducting loops on chip to be used to provide flux biasing for superconducting circuits, like qubits. We developed three types of flux memory and draw comparisons among them for circuit design. We demonstrate the utility of flux memory by using an in-situ flux detector and characterize each approach and further demonstrate that once flux is set in a memory cell, benchtop DC control sources can be powered off, leaving the on-chip flux bias in place. We propose that flux memory can be arranged in a two-dimensional configuration to multiplex control signals and reduce how line counts scale ($N^2\ devices \rightarrow 2N$ control lines), and our experimental results pave the path to the proposed scalability. We demonstrate the use of flux memory to flux bias a transmon qubit and show the tunability of the qubit state to a target frequency which remained stable on chip for 20 hours.

## Introduction

Significant progress has been made in recent years in the development of integrated circuits that use superconducting materials and devices for a wide variety of computing applications [1], [2], [3], [4], [5], [6], [7], [8], [9], [10]. Superconducting components often need to be biased to a working configuration by current, flux, or combination of the two to operate. Bunyk et al [11] pioneered the concept of flux memory (Fig. 1a) by developing a flux digital-to-analog converter (DAC) to control flux qubits with zero static power dissipation. Since then, a number of works have focused on various aspects related to their own applications [12], [13], [14], [15]. We developed several flux memory circuits that we categorize into three types: a single Josephson junction (SJJ) also known as an rf-SQUID, a flux lock that requires 2 or more Josephson junctions (JJ), and cryotron flux memory. All three of the flux memory cells are designed to apply persistent flux to another superconducting circuit, such as a transmon or a flux qubit. The experimental results are collected below 30mK to verify our design.

## Design and Methods

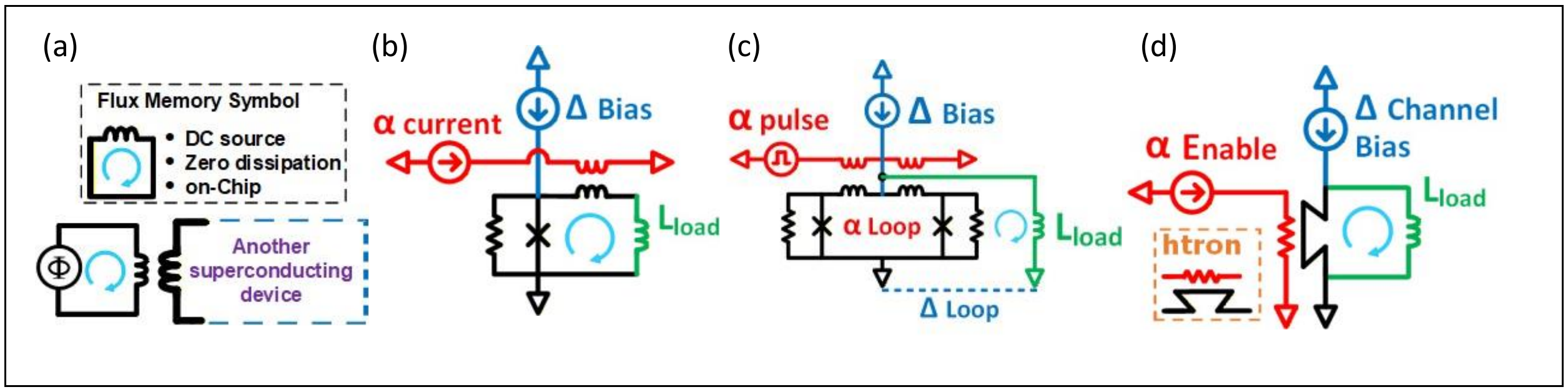

Fig. 1. Flux memory technology and schematics of the three types of flux memory. (a) Flux memory cell stores flux as memory and can provide flux bias to another superconducting device. (b) The SJJ, also known as RF SQUID, (c) Flux Lock with 2 or more JJs (d) Cryotron activated flux memory. Inset: an orange dash-lined box indicates the symbol for an 'htron' consisting a nearby resistive heater and a constriction.

In this work, we developed three types of circuit elements that store flux. A general circuit schematic is depicted in Fig. 1a, with each variant depicted in the rest of Fig. 1. Each of the three flux memory approaches has its own trade-offs depending on their use case. For a particular set of requirements, a designer may adopt any combination of three types to meet requirements for a given application. The common theme shared among all three types is that they are designed to serve as a direct current (DC) source (equivalently DC flux bias via a mutual inductor), to be programmable, use ultra-low power, and to be compatible with on-chip integrated circuit (IC) and monolithic microwave integrated circuit (MMIC) designs. Each of the flux memory designs consume zero power when they are not being programmed, dissipate zero heat while storing flux, and provide DC flux to another superconducting device with steady bias for hours or longer.

The first cell type, the single Josephson junction (SJJ) memory, is a hysteretic RF SQUID and shown in Fig. 1b. It uses a loop with a single JJ coupled to the current bias lines through a mutual inductance, or a mix of galvanic connections and mutual inductance. We adopt a resistively shunted JJ to ensure that we reliably pump a single flux quantum (SFQ) into the inductive load. The advantages of SJJ are that it is the simplest design, with fast tune-up while achieving a reduction in bias lines for large arrays. The flux storage is limited by absolute critical current so higher values of stored SFQ become inaccessible, and it needs a large mutual inductance for flux drives.

The second cell type, the flux Lock, utilizes two resistively shunted JJs in a loop with both a galvanic biases and biases coupled to the loop by mutual induction similar to what has been done previously in literature [11], as shown in in Fig. 1c. Two of them can be employed for fine and coarse load biases and they are less sensitive to parameter inaccuracies. The main advantage of this approach is digital control with a clear one-to-one correspondence between an input pulse and a single flux quantum loading event. Compared to the SJJ it also requires less alpha-flux, and therefore less current or a smaller mutual inductor (and hence footprint).

The third cell type, the heater cryotron (htron) [12], utilizes a JJ-free superconducting loop biased by a galvanically coupled control line, as shown in in Fig. 1d, flux can be threaded or expelled in the loop by injecting current into the $\alpha$ enable line, powering the heater, generating a local hot spot inside a constriction in the superconducting bias channel arm, driving part of the superconducting path normal. Magnetic fields can enter or exit via this normal metal region to trap flux in the loop. The advantage of this approach is that current compliance is limited by the superconducting loop, not the critical current of a Josephson junction, so all possible trapped flux states are accessible, but it necessarily dissipates heat in the cold space for this mode of operation which could become a concern in cooling power constrained environments.

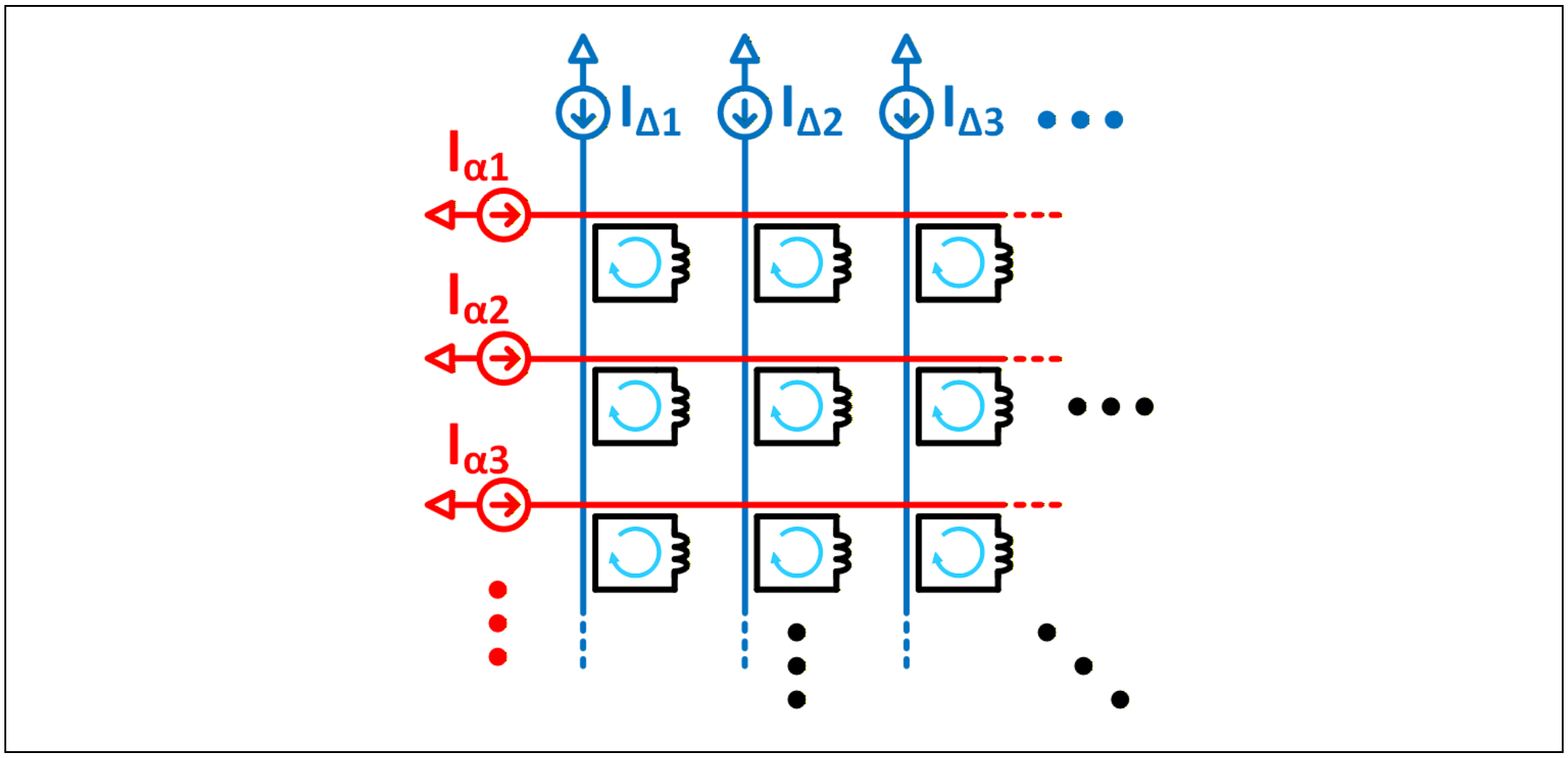

Fig. 2. A schematic of the proposed 2D addressing scheme with horizontal and vertical control lines

We designed the flux memory circuits to control or calibrate a large number of superconducting circuits. The three types of flux memory are developed to be controlled by two input lines $\alpha$ & $\Delta$, as show in Fig. 1b-d, resembling word and bit lines in digital memory. We laid out the flux memory cells in a 2-dimensional (2D) grid with $\alpha$ rows and $\Delta$ columns of control lines, as shown in Fig. 2. By selectively biasing the $\alpha$ and $\Delta$ control lines, we can control any cell in the 2D grid, effectively reducing the number of DC bias lines from $N^2$ for individually addressed devices to 2N for the sides of the grid. Furthermore, signals to $\alpha$ and $\Delta$ control lines can be provided by fewer lines from room temperature using a MUX. We send enable and/or dis-enable signals across the $\alpha$ horizontal lines to select the rows to be addressed, and then send signals down the $\Delta$ vertical lines to program the memory array.

| Memory type | Nominal step size ($\boldsymbol{m\Phi_0}$) in design for $\Phi_{\text{device}}$ | Experimental step size ($\boldsymbol{m\Phi_0}$) in test for $\Phi_{\text{device}}$ |
|---|---|---|
| Single JJ | 5.2 | 4.63 |
| Flux lock | 22.5 | 20.8 |
| Htron | 62.5 | 43.8 |

Table 1. Flux memory design values and experimental results. We design a nominal step size to meet different flux memory resolution requirements for different targeted devices. The flux memory circuits are laid out after performing inductance simulations to ensure wide working parameter margins to overcome foundry process variation. We compare experimentally extracted step sizes of each memory type and achieve within 10% targeting goal for JJ based memory and 30% for cryotron-based design. The current simulation does not include modeling of kinetic inductance [12, 15].

The flux memory components were designed to bias various superconducting loads connected to the flux memory output. Fundamentally, the smallest memory step in any flux memory cell is $\Phi_0 \approx 2.07 \times 10^{-15}\ Wb,$ the magnetic flux quanta, with step size of the flux applied to a target device being set by the designed mutual and load inductances. The operating principle of flux memory cells can be derived from some basic facts about superconductivity. For a wire to remain superconducting the current cannot exceed the critical current $I < I_c$, and $IL = N\Phi_0$. The inductance of a Josephson junction (JJ) can be obtained from $V = L\dot{I}$ and $I = I_c \sin\phi$ so that $L_{JJ} = \frac{\Phi_0}{2\pi\, I_c \cos\phi}$. Consider a loop with a single JJ and an inductor. The requirement $I < I_c$, becomes $\frac{\Phi}{L} < \frac{\Phi_0}{2\pi\, L_{JJ}}$. With the definition, $\beta = \frac{L}{L_{JJ}}$,

$$N < \frac{\beta}{2\pi}$$

Higher $\beta$, leads to higher values for N, the number of stored flux quanta and the value of $\beta$ is directly proportional to $L_{loop}$ so increasing $\beta$ requires larger loops. A tradeoff in storing more flux quanta is the depth of potential energy minima which get shallower in response to larger N approximated by the expression, $\Delta E_N \approx \frac{2\Phi_0 I_c}{2\pi}\left(1 - \frac{N\Phi_0}{I_c}\right)^{\frac{3}{2}}$ [Tinkham]. The main escape mechanisms are thermal activation, which has contributions from control line noise, for higher temperatures and macroscopic quantum tunneling in the zero-temperature limit. The desire for a small physical footprint for the inductors and the stability of flux stored with shallower potential wells as N increases sets the limit for the flux storage cells. The range and step size of the flux output of one memory cell is determined by the following equation:

$$\Phi_{max} = \mathrm{N} \cdot \text{step size} = \mathrm{N}\frac{\mathrm{M_c}}{\mathrm{L_{load}}}\Phi_0 \tag{1}$$

where N is number of allowed memory states in one polarity, $\mathrm{L_{load}}$ is the inductance of the loop storing the flux, $\mathrm{M_c}$ is the mutual inductance between the cell and the flux biased device. The flux memory can be designed to have multiple cells in two stages to cover coarse and fine steps to one load device so that the bias can be tuned with small, $\Phi_{max}^{small}$, and large, $\Phi_{max}^{large}$, step sizes. For two staged cells, the coarse memory cell provides the flux bias range, and the fine memory cell provides the resolution. Table 1 shows comparison of as designed step size and experimentally extracted step size for each of the memory types, where we denote $\Phi_{\mathrm{device}}$ as flux received by the device originating from flux memory. Fabrication variation of the inductance in a given metal layer accounts for the differences.

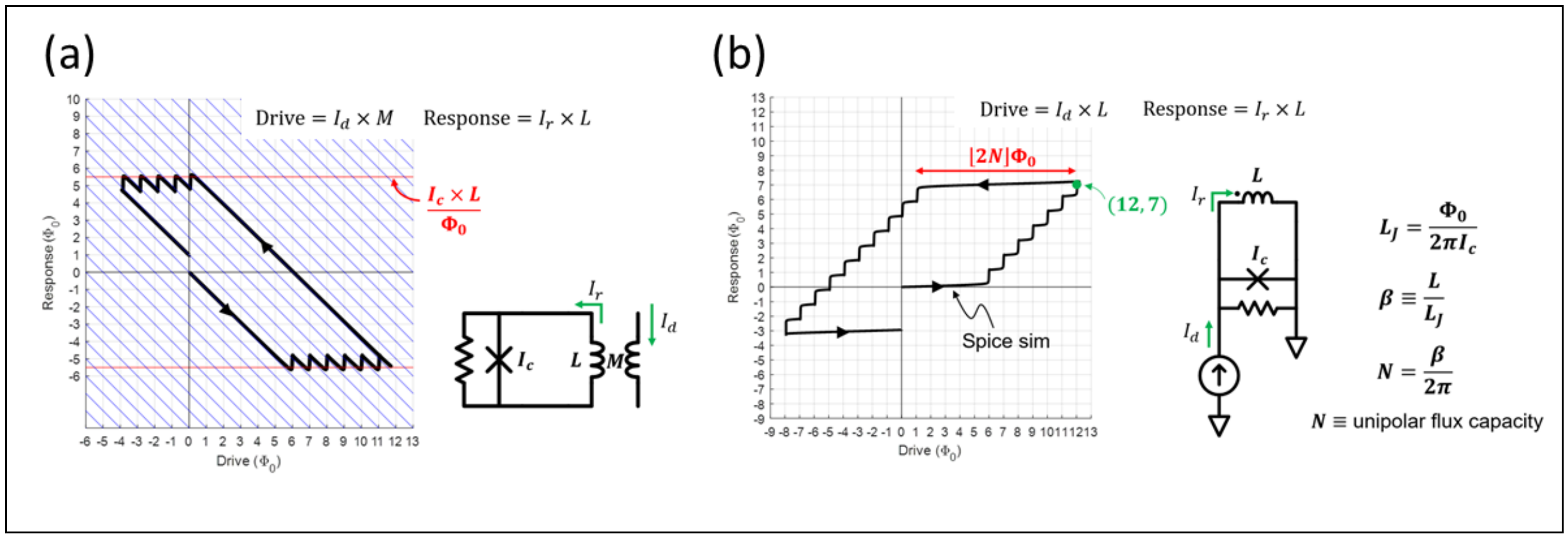

Fig. 3. An SJJ flux memory SPICE simulation. The circuit schematic and key equations used in simulation are shown on the right of each panel for two SJJ circuit variants: (a) Flux drive with a mutual connection between the cell and control current. The rails indicating maximum memory value are drawn as horizontal lines and annotated in red. (b) Flux drive with a galvanic connection between the cell and control current. An allowed drive-response example contour is traced out as the black thickened line for a possible programmable memory operation. The maximum memory value annotated in red. Two allowed drive-response example contours for (a) and (b) are traced out as black thickened lines for a possible programmable memory operation with arrows indicating direction of the programing sequence.

We use a SPICE simulation model to capture the behavior of the SJJ flux memory, as shown in Fig. 3. The $\alpha$ and $\Delta$ modulation shown in Fig. 1b can be broken down independently into sub-circuits consisting of 1 JJ with their respective input current connection. We define the 'drive' flux as the flux we are injecting into the cell via either mutual or galvanic connections, and we define the 'response' current as the counter flux the loop produces. The sawtooth pattern shown in Fig. 3a, and the step pattern shown in Fig. 3b, occur when the maximum amount of flux is stored within the loop. In test, a sequence of steps is programmed to take a walk along the response-drive curve to verify our design. We characterized the memory experimentally, which allows us to convert the drive and response currents to the flux quanta stored in the loop $N\Phi_0$.

The htron flux memory is controlled with a galvanically connected current bias line ($I_\Delta$), and an alpha line current ($I_\alpha$) that passes through a resistor embedded near the superconducting loop, as shown in Fig. 1d. The cell state is programmed by first setting $I_\alpha$ high enough to power the heater and drive part of the superconducting loop normal. A current is then sent through $I_\Delta$ that is equal to the desired persistent current in the loop. $I_\alpha$ is then turned off to make the loop superconducting again, before which $I_\Delta$ is turned off. The $\Delta$ current is related to the desired input flux ($\Phi_{device}$) by:

$$I_{\Delta,CH} = \frac{\Phi_{device}}{M_c} \tag{2}$$

## Experimental Results

### Flux Detection with a Tunable Resonator

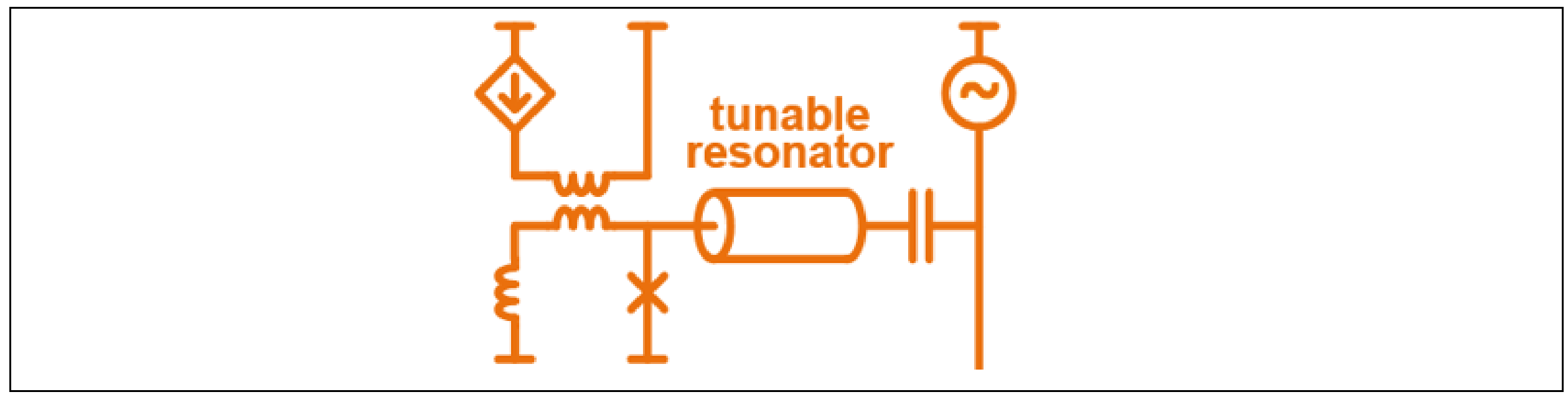

Fig. 4. Schematic of the tunable resonator flux detector. There is an independent bias line to verify that the tunable resonator yielded and functions above the SQUID, otherwise the SQUID bias comes from the flux memory through a mutual inductor. The SQUID can be thought of as a tunable inductance, thus shifting the resonant frequency of the LC-circuit comprised of a SQUID and resonator when flux is applied to its loop.

The flux signal is read out by interrogating a tunable resonator with a probe tone, as depicted in Fig. 4. The tunable resonator's frequency tracks the amplitude of the set current. An important feature of the flux detector is that it is periodic in the flux applied to the SQUID loop, so as shown in the data throughout this paper, the resonance will return as a function of the bias currents.

#### 1) SJJ Flux Memory Results

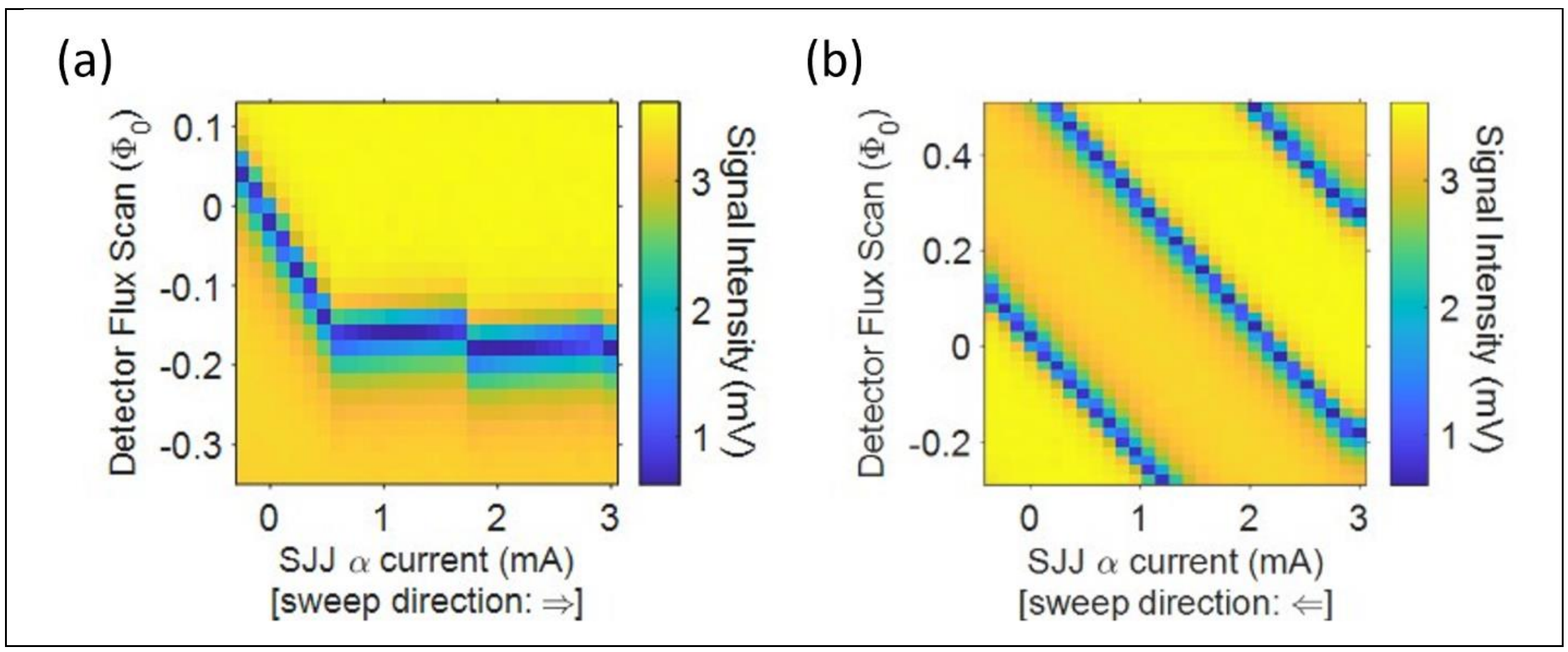

Fig. 5. Experimental demonstration of a single SJJ flux memory cell operation. We use a flux detector to monitor the flux at the memory cell output. (a) Forward sweep of a current into the $\alpha$ control line. The data indicates the initial flux state at the 0 mA $\alpha$-current. (b) "Backwards" sweep along the x-axis from high $\alpha$-current to low along the same parameter range and steps as in (a). The data indicates the final flux state of the cell at the 0 mA $\alpha$ current after completing the hysteretic loop in parameter sweep. Repeated signals are an artifact periodicity of the flux detector.

The SJJ memory cell is the simplest, so as a first step to understanding SJJ flux memory, we use an in-situ flux detector to probe a single memory cell in comparison with simulations. The flux detector scans values in a programmable range and step sizes while generating a measurable signal at each sampled step. When the scanned value matches the actual received flux from the memory cell, the detector signal intensity diminishes, as shown in Fig. 5. We scan the flux detector value along the y-axis and sweep a flux memory control parameter in the x-axis while reading out detector signal to investigate into the memory cell circuit.

We observe the expected hysteretic behavior of a single memory cell by sweeping $\alpha$ current in x-axis with the same sweep range and steps but in the opposite direction, as shown in Fig. 3. We attribute the hysteretic behavior of the memory cell to the stored flux. We perform a reset protocol to erase the memory. As a result, the detector tracks the relative flux starting near 0 $\Phi_0$ at 0 mA $\alpha$ current, and then decreases the drive until the detector reads just above $-0.2\,\Phi_0$, sweeping $\alpha$ from 0 mA to 3 mA, as shown in Fig. 5a. We then sweep $\alpha$ down from 3 mA to 0 mA, as shown in Fig. 5b, and we found that the detector measures 0.4 $\Phi_0$ at 0 mA $\alpha$ current. Our results agree with hysteretic behavior as a function of the $\alpha$ current that our model predicts, as shown in Fig. 3.

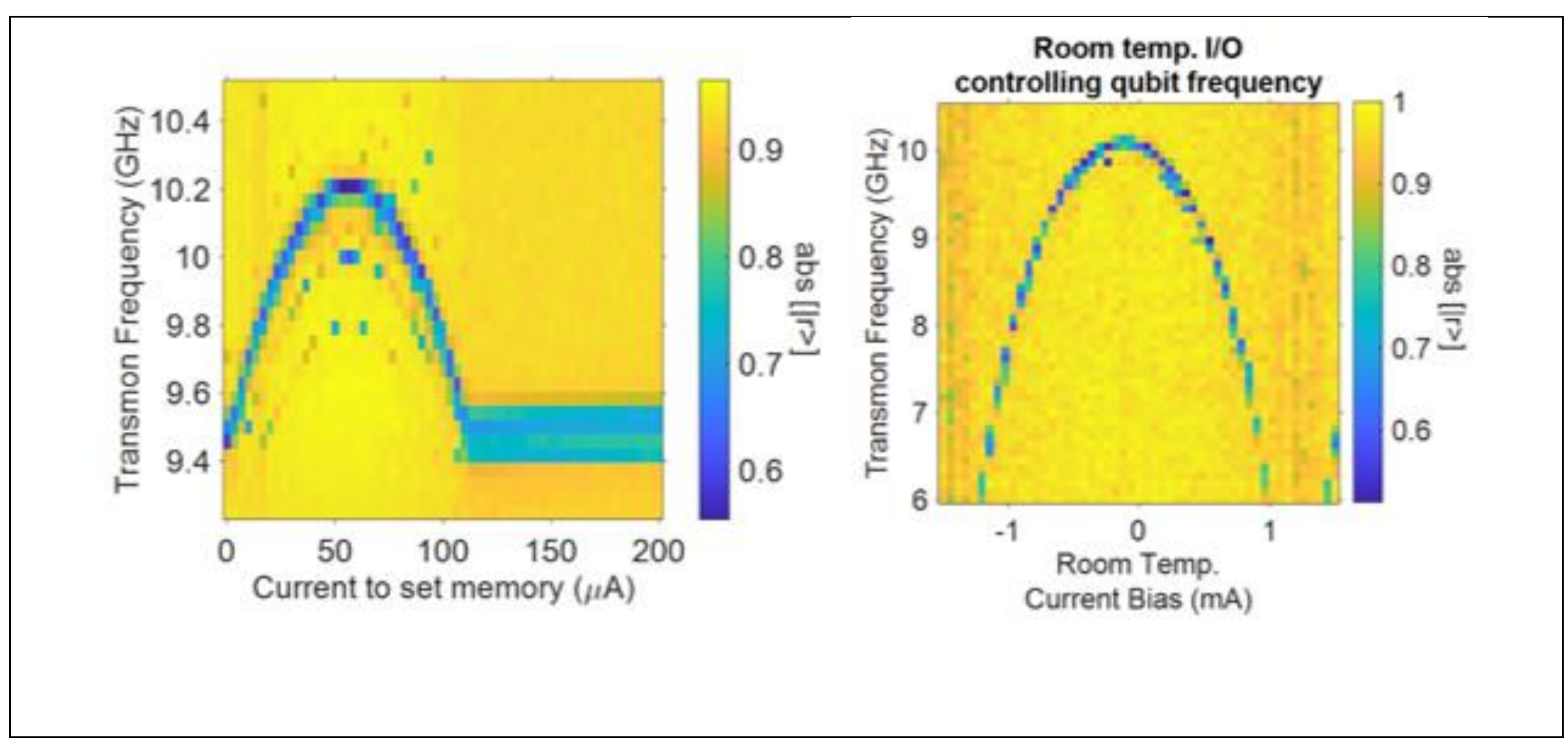

Fig. 6. Experimental demonstration of a single SJJ flux memory cell operation. We use an SJJ memory circuit to provide DC flux bias to a transmon qubit. We compare this with room temperature control. The flat region of the transmon frequency is when the SJJ circuit hits compliance set by the critical current of the junction and cannot provide more flux.

After verifying flux tunability, we used the SJJ flux memory circuit to provide DC bias to a flux tunable transmon with an upper sweet spot of 10.2 GHz, shown in Fig. 6. We verified the behavior using a current source at room temperature. The transmon frequency is approximately proportional to $\sqrt{|\cos(\pi\Phi_{\mathrm{device}}/\Phi_0)|}$

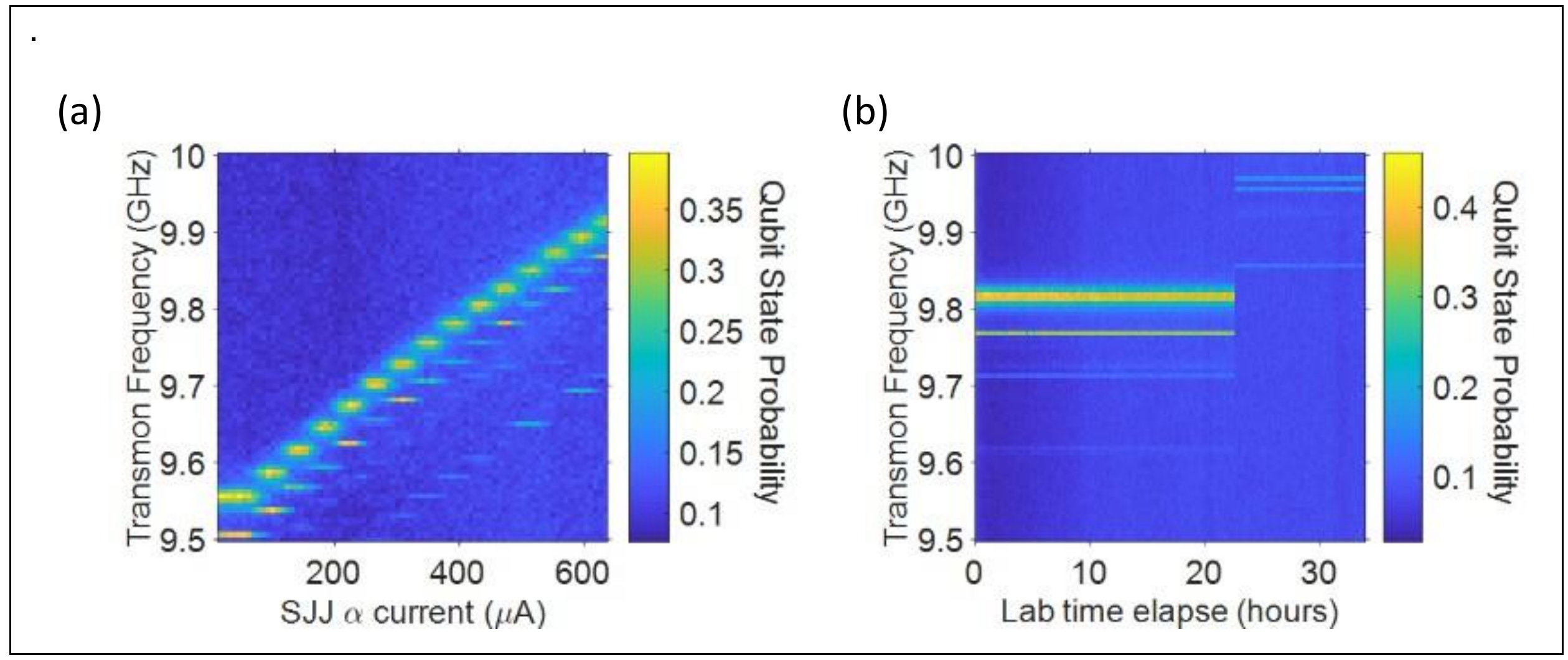

Fig. 7. Demonstration of the SJJ flux memory biasing a transmon. a) The present memory control displays a single memory state step change shifting the transmon frequency by 30MHz. The narrower steps are results of power leakage from the continuous wave (CW) microwave IQ mixing

side band. b) We test the stability of the memory state by tuning the qubit frequency to 9.8 GHz and monitoring this resonance as a function of time. Over 20 hours, using 1 minute measurement intervals, the resonance frequency remains fixed, and we observe no jumps in frequency.

To connect the single-flux quanta events to the bias provided to the transmon, we loaded a finite amount of flux into the memory cell by use the $\Delta$ current to tune the frequency of the transmon from approximately 9.6 GHz to just under 10 GHz, as shown in Fig. 7a. We increased the $\alpha$ current to unload flux from the cell while we measure the transmon with a microwave readout scheme. This readout scheme allowed us to see the discrete SFQ steps in $\Phi_{\text{device}}$ of the flux biased transmon device.

We tested the stability of the flux memory by storing flux in the cell, tuning the transmon frequency to near 9.8 GHz, and shutting down all room temperature power supplies powering the flux memory. Once biased, the flux memory did not consume power or dissipate energy. We found the transmon remained biased with a constant flux bias for over 20 hours, shown in Fig. 7b, after which the transmon frequency returned to its zero-flux resonance frequency of 10.2 GHz, outside the range shown in Fig. 7b. The flux memory state eventually experienced a flux event, likely due to thermal noise [16]. The lifetime of the state as more flux is stored is determined by the decreasing potential energy barrier between the flux states. This demonstrates that SJJ flux memory can be used as a current source for biasing superconducting components on a time scale of the order of a day.

#### 2) Flux Lock Results

The next approach is dubbed the 'flux lock,' shown in Fig. 1c. The flux lock operates as follows: There is an overall bias current setting the compliance. Each enable pulse pushes a $\Phi_0$ into the loop, which shifts one of the JJs phases by $2\pi$. One stored flux quantum results in a bias of $\frac{\Phi_0}{2}$ for the output inductor. As the injected flux is removed, the other JJ triggers which loads another $\frac{\Phi_0}{2}$ into the output inductor. Flux is 'locked' into the output inductor with each pulse from the input line, via the DC-SQUID injector.

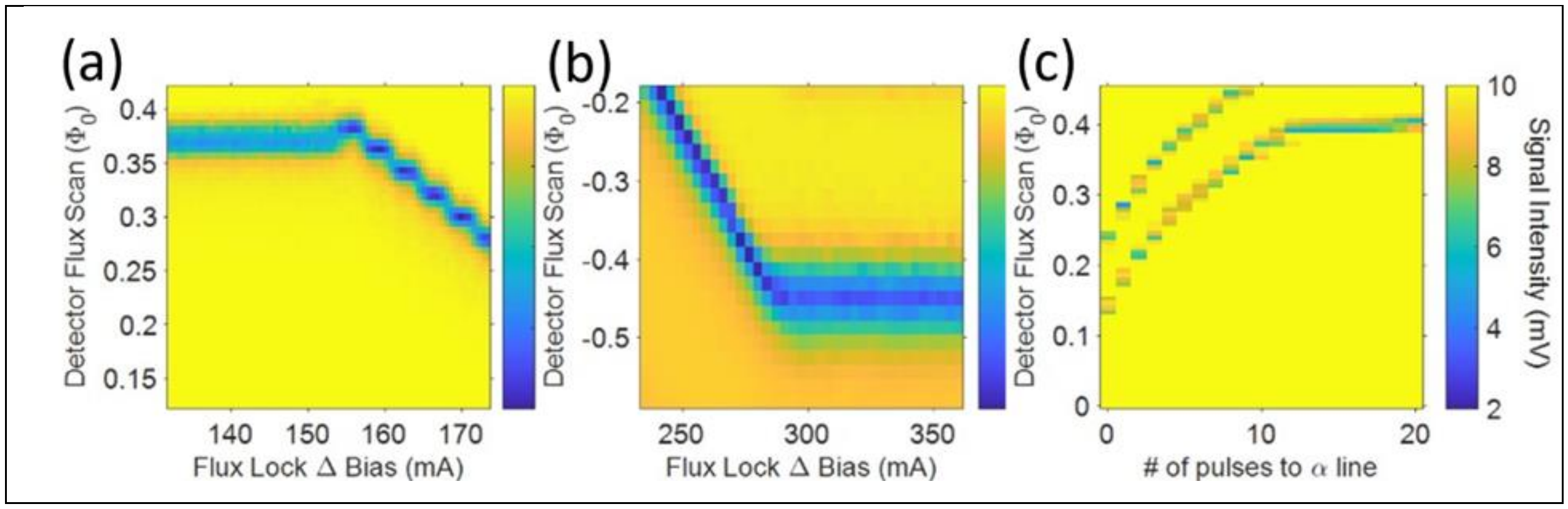

Fig. 8. Experimental demonstration of a single flux lock memory cell operation. We sweep the bias current in Δ line to lock in flux and observe single flux quanta (SFQ) modulation in the flux signal at the detector indicating that the memory state is resolving SFQ events.

First, we tested the flux lock memory cell to our tunable resonator flux detector so that we could measure its performance and compare it to simulations. We found we can step the output of a single flux lock cell by addressing its memory state via the two inputs, as shown in Fig. 1c. We increased the Δ bias (refer to Fig. 1) to modulate the flux, as shown in Fig. 8a, and found that at first the Δ current passes through the DC SQUID rather than the load, until the flux stored in the Δ loop exceeded the stored current threshold of 86% of $2I_c$. Current noise of order $I_c$ can lower the injection and saturation thresholds. We measured the stored flux quanta levels in a manner similar to how we measured them in the SJJ flux memory experiment to confirm that we can control the resolution with the minimum designed memory step. We measured the maximum amount of flux that we can store in the cell by measuring when the flux output reached saturation. The maximum capacity of the memory is scales with the critical current of the JJ's used in the cells (N < $LI_c/\Phi_0$). After we had confidence that the Δ input bias can store flux in the memory cell, we stored a finite flux in the cell to set an initial memory state and sent current pulses into the $\alpha$ line to "unlock" the flux lock, emptying the memory cell, as shown in Fig. 8c. Each pulse we sent down the $\alpha$ line set the cell into a distinct memory state. We found that we needed $13$ pulses to reset the cell, as shown by the plateau in Fig. 8c. We found that once the cell reached an empty state, sending more pulses down the $\alpha$ line had no effect on the output, as is expected since there was no longer any trapped flux.

### 3) HTron Results

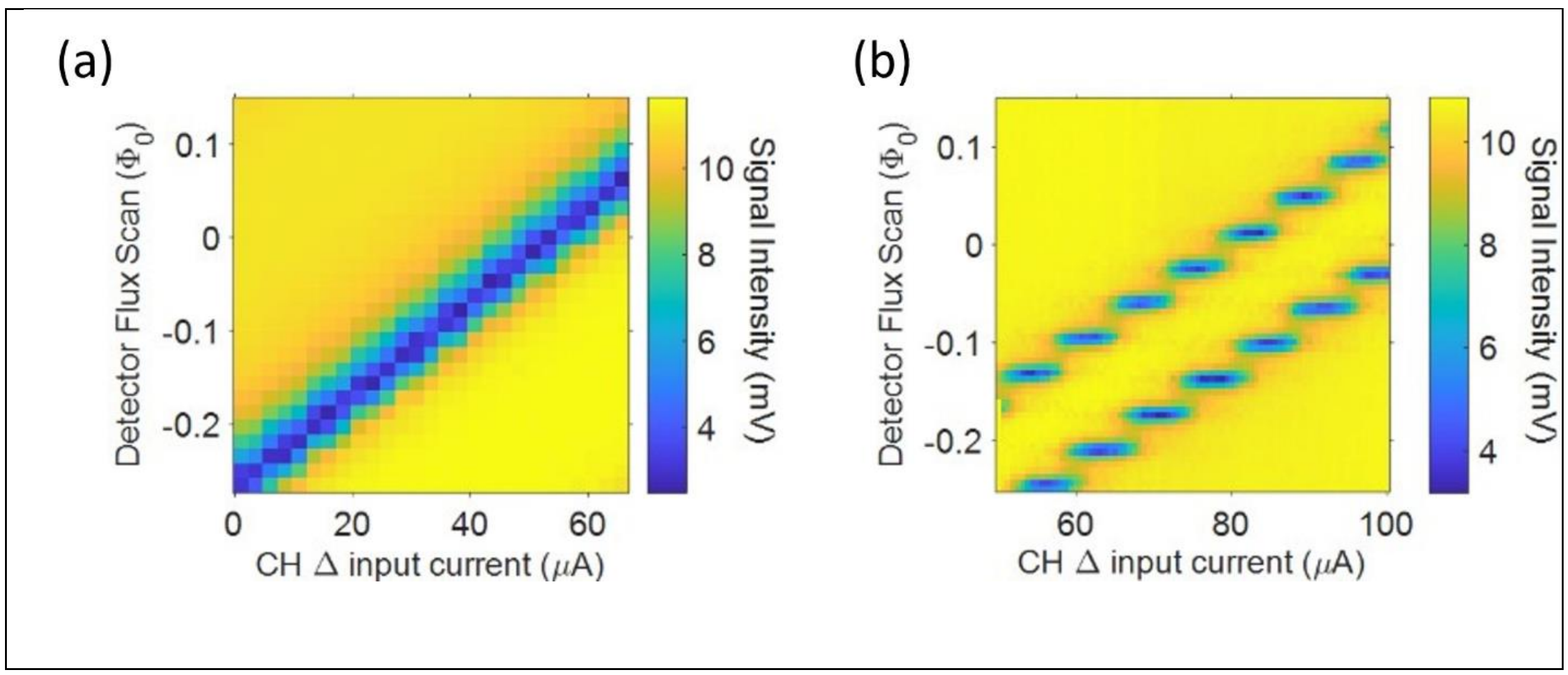

Fig. 9. Experimental demonstration of an htron flux memory cell. (a) After tuning to a negative initial value, we observe a linear dependence of flux modulation with increasing channel input current (CH Δ) into the htron. (b) Increased resolution of the flux detector to resolve single flux quanta control of the htron cell and thereby confirming that the memory state.

Our third approach used a direct galvanic control of an 'htron,' (heater cryotron). We tested the novel htron flux memory and compared the experimental results to simulations. We found that we can program the htron to output a wide range of flux bias currents to bias a superconducting load, as shown in Fig. 9a. The desired output flux will be set with the bias current. The flux biasing and detection of the current into the load is, again, characterized by a flux detector, as shown in Fig. 9b. The detector readout signal was recorded while the channel input current (CH Δ) is swept. Each pixel of the color plots represents a sequentially activated htron, and the CH Δ input current is the current directed towards the load device and is stored as a circulating persistent current associated with a flux value. Our results show the expected linear dependence, given by $N\Phi_{\mathrm{o}} = L_{load}\, I_{\Delta}$. We verified that the fundamental step size of the htron memory cell was in units of flux quantum ($\Phi_o$), scaled by the inductance of the circuit, by using a sufficiently high resolution to observe individual steps in the flux bias while sweeping CH Δ, as shown in Fig. 9b. The step pattern confirms the minimum memory step size and verifies the design values of htron flux memory. Known memory states with predictable spacing between memory steps can be obtained with this control. The advantage of the htron is that it does not use JJs, so the range of flux quanta that can be stored in memory in the htron is not limited by JJ critical current and therefore much larger than the other two flux memory types reported in this paper. The htron is capable of storing hundreds of flux quanta because it is only limited by the current of the traces used in the superconducting loop.

## Conclusion

We have designed three types of superconducting memory circuits: the single-JJ flux memory, the two-JJ flux memory, and the htron flux memory, which all operate at cryogenic temperatures. The SJJ results verified the proof of principle and the stability of flux memory. The flux lock approach shows increased tunability and unloading of flux was demonstrated (for reprogramming). The htron has the distinct advantage of being able to store more flux than the other two approaches. In addition to flux storage, these circuits elements can also be used for controlling other superconducting circuits that are mutually coupled to these circuits. All three of these circuit types can be programmed via two independent control signals, paving the way for the addressability of these components in a proposed 2D array scheme which reduces DC bias of $N^2$ devices to only 2N lines with further reductions possible by multiplexing. The number of control lines scales linearly number of devices and once programmed, they dissipate no power. We have also shown the stability of these non-volatile memory elements as function of time in the order of 20 hours which makes them suitable candidates for quantum information processing applications.

### *Authors (in alphabetical order by last name):

Gregory R. Boyd, Jeremy Clark, Mark Covington, David G. Ferguson, Aref Fouladi, Keith D. Hillaire, Moe Khalil, John McFarland, Aaron Pesetski, Anthony J. Przybysz, John X. Przybysz, Sambarta Rakshit, Aruna N. Ramanayaka, Brian Sears, Robert Smith, Colin Stack, Robert M. Young, Tony X. Zhou

## References

[1] D. A. Abanin et al., “Observation of constructive interference at the edge of quantum ergodicity,” Nature, vol. 646, no. 8086, pp. 825–830, Oct. 2025, doi: 10.1038/s41586-025-09526-6.

[2] Q. P. Herr, A. Y. Herr, O. T. Oberg, and A. G. Ioannidis, “Ultra-low-power superconductor logic,” J. Appl. Phys., vol. 109, no. 10, p. 103903, May 2011, doi: 10.1063/1.3585849.

[3] S. Bravyi, A. W. Cross, J. M. Gambetta, D. Maslov, P. Rall, and T. J. Yoder, “High-threshold and low-overhead fault-tolerant quantum memory,” Nature, vol. 627, no. 8005, pp. 778–782, Mar. 2024, doi: 10.1038/s41586-024-07107-7.

[4] S. Bravyi, O. Dial, J. M. Gambetta, D. Gil, and Z. Nazario, “The future of quantum computing with superconducting qubits,” J. Appl. Phys., vol. 132, no. 16, p. 160902, Oct. 2022, doi: 10.1063/5.0082975.

[5] J. Yang et al., “High-Temporal-Resolution Measurements of the Impacts of Ionizing Radiation on Superconducting Qubits,” Feb. 26, 2026, arXiv: arXiv:2602.23544. doi: 10.48550/arXiv.2602.23544.

[6] J. A. Strong et al., “A resonant metamaterial clock distribution network for superconducting logic,” Nat. Electron., vol. 5, no. 3, Art. no. 3, Mar. 2022, doi: 10.1038/s41928-022-00729-7.

[7] J. A. Grover et al., “Fast, Lifetime-Preserving Readout for High-Coherence Quantum Annealers,” PRX Quantum, vol. 1, no. 2, p. 020314, Nov. 2020, doi: 10.1103/PRXQuantum.1.020314.

[8] A. L. Graninger et al., “Superconducting on-chip solenoid for Josephson junction characterization,” Appl. Phys. Lett., vol. 115, no. 3, p. 032601, Jul. 2019, doi: 10.1063/1.5110170.

[9] A. L. Graninger et al., “Critical Current Modulation in Josephson Junctions Contacted by Redundant Vias,” IEEE Trans. Appl. Supercond., vol. 32, no. 1, pp. 1–5, Jan. 2022, doi: 10.1109/TASC.2021.3128626.

[10] A. L. Graninger et al., “Microwave Switch Architecture for Superconducting Integrated Circuits Using Magnetic Field-Tunable Josephson Junctions,” IEEE Trans. Appl. Supercond., vol. 33, no. 6, pp. 1–5, Sep. 2023, doi: 10.1109/TASC.2023.3268547.

[11] P. I. Bunyk et al., “Architectural Considerations in the Design of a Superconducting Quantum Annealing Processor,” IEEE Trans. Appl. Supercond., vol. 24, no. 4, pp. 1–10, Aug. 2014, doi: 10.1109/TASC.2014.2318294.

[12] B. A. Butters, R. Baghdadi, M. Onen, E. A. Toomey, O. Medeiros, and K. K. Berggren, “A scalable superconducting nanowire memory cell and preliminary array test,” Supercond. Sci. Technol., vol. 34, no. 3, p. 035003, Jan. 2021, doi: 10.1088/1361-6668/abd14e.

[13] L. Jiang, “On-chip direct-current source for scalable superconducting quantum computing,” Phys. Rev. Appl., vol. 24, no. 3, 2025, doi: 10.1103/qwvv-mz8s.

[14] P. Szypryt et al., “Kinetic inductance current sensor for visible to near-infrared wavelength transition-edge sensor readout,” Commun. Eng., vol. 3, no. 1, p. 160, Nov. 2024, doi: 10.1038/s44172-024-00308-y.

[15] M. Onen, M. Turchetti, B. A. Butters, M. R. Bionta, P. D. Keathley, and K. K. Berggren, “Single-Photon Single-Flux Coupled Detectors,” Nano Lett., vol. 20, no. 1, pp. 664–668, Jan. 2020, doi: 10.1021/acs.nanolett.9b04440.

[16] Fulton, T. A. and Dunkleberger, L. N., “Lifetime of zero-voltage state in Josephson tunnel junctions,” PhysRevB, vol. 9, no. 11, pp. 4760-4768, Jan. 2020, doi: 10.1103/PhesRevB.9.4760